\documentstyle[prl,preprint,aps]{revtex}
\def\2nu{$\tilde\chi$}
\def\neut{$\tilde\chi$ \,}
\def\chN2{$\tilde\chi N$}
\def\chiN{$\tilde\chi N$ \,}
\def\131x{$^{131}$Xe}
\def\i127{$^{127}$I}
\def\x129{$^{129}$Xe}
\def\t125{$^{125}$Te}
\def\n23{$^{23}$Na}
\def\sp{${\bf S}_p \,$}
\def\sn{${\bf S}_n \,$}
\begin{document}
\draft
\tighten
\preprint{ }
\title{Spin-Dependent Neutralino-Nucleus Scattering for 
$A \sim 127$ Nuclei}
\author{M.T. Ressell}
\address{W.K. Kellogg Radiation Laboratory, 106-38, California 
Institute of Technology, Pasadena, CA 91125}
\author{D.J. Dean}
\address{Physics Division, Oak Ridge National Laboratory,
Oak Ridge, TN 37831}
\date{\today}
\maketitle
\begin{abstract}

We perform nuclear shell model calculations of the neutralino-nucleus
cross section for several nuclei in the $A = 127$ region. Each of the
four nuclei considered is a primary target in a direct dark matter
detection experiment.  The calculations are valid  for all relevant values
of the momentum transfer.  Our calculations are performed in the 
$3s 2d 1g_{7/2} 1h_{11/2}$ model space using extremely large
bases, allowing us to include all relevant correlations.  We also
study the dependence of the nuclear response upon the assumed nuclear
Hamiltonian and find it to be small.  We find good agreement with 
the observed magnetic moment as well as other obervables for  the 
four nuclei considered: $^{127}$I, $^{129,131}$Xe, and
$^{125}$Te.

\end{abstract}
 
\section{Introduction}

\narrowtext
 
An ever increasing amount of evidence indicates the existence of
large amounts of dark matter in the Universe \cite{r:jkg}.  
Despite this overwhelming evidence,
the exact nature of the dark matter remains a mystery.  Numerous candidates
have been proposed, including both baryonic and nonbaryonic
matter \cite{r:dmrev}.  Observations reveal that some of the dark
matter in the Galactic halo is baryonic, consisting of MACHO's
\cite{r:macho}; however, present data indicates that MACHO's 
cannot account for all of the dark matter implied by the
Galactic rotation curve \cite{r:gates}.  Furthermore, a number of
arguments based upon large scale motions in the Universe and large
scale structure formation indicate that $\Omega \approx 1$, which is far
in excess of the bounds on $\Omega_{\rm Baryon}h^2 \leq 0.026$ 
arising from cosmic nucleosynthesis \cite{r:nucsyn}.
All considerations point toward non-baryonic matter comprising
a sizable fraction of the Universal density.  If this is true, what is
the dark matter?

Among the best motivated, and hence highly favored, of the nonbaryonic
dark matter candidates is the lightest supersymmetric particle (LSP).
Experimental and theoretical considerations indicate that the LSP
is a neutralino, $\tilde\chi$, consisting of a linear combination
of the supersymmetric partners of the photon ($\tilde\gamma$), 
the Z ($\tilde Z$), and 2 Higgs bosons ($\tilde H_1$ and $\tilde H_2$).
Note that the $\tilde\gamma$ and $\tilde Z$ are themselves linear 
combinations of the supersymmetric partners of 
the neutral W ($\tilde W_3$) and
B ($\tilde B$) bosons, hence the neutralino composition is typically
written as
\begin{equation}
{\tilde \chi} = Z_1 {\tilde B} + Z_2 {\tilde W_3} + Z_3
{\tilde H_1} + Z_4 {\tilde H_2}. 
\end{equation}
The motivation for
supersymmetry (SUSY) arises naturally in modern theories of
particle physics \cite{r:jkg,r:hk}, although the \2nu's potential as a dark
matter candidate was not realized until later \cite{r:dm1st}. For
a very large region of SUSY parameter space, neutralinos provide
densities that are in accord with the measured value of 
$\Omega$, thus explaining the dark matter.  The \neut is also
detectable in at least two ways: indirectly, through the products of
$\overline{\tilde\chi}$\neut annihilation in the Sun, Earth, and
Galactic halo \cite{r:jkg,r:marc}, or directly, via elastic (and
inelastic \cite{r:japan}) neutralino-nucleus (\chN2) scattering in
a detector \cite{r:jkg,r:gw}.  In either case, the elastic
\chiN scattering cross section is an essential ingredient.
In this paper we discuss nuclear structure calculations relevant to
\chiN scattering for several nuclei which are primary constituents
of many current and planned direct detection experiments
\cite{r:japan,r:bottino,r:italy,r:boulby,r:bernabei,r:cline}.

Physics at three distinct energy scales governs \chiN scattering.
The composition and mass of the \2nu, and hence its interaction with
quarks are fixed near the electroweak scale.  The interaction of 
neutralinos with protons and neutrons is determined by the quark
distribution (both spin and density) within the nucleon, which 
is determined at the QCD scale.  At the modest momentum transfers
available to dark matter neutralinos the \neut interacts with the
entire nucleus, not individual nucleons within it.  Thus,
nuclear structure plays an important role in determining the
\chiN cross section.  The uncertainties in the electroweak scale
physics (the SUSY part of the problem) are typically handled by considering
large sweeps through SUSY parameter space \cite{r:jkg,r:bednayakov}.  The
QCD scale physics is currently the focus of much study and the
relevant nucleon matrix elements continue to be measured 
with high precision
\cite{r:spin}.  The necessary nuclear physics is not measurable for
most nuclei but is amenable to calculation through a variety of methods.
Here we apply the nuclear shell model to the nuclei
\i127, \x129, \131x, \t125, and \n23 (this last nucleus we discuss
in an appendix) in order to provide a consistent
and correct set of nuclear input physics for determining the
\chiN cross section.

The  \chiN scattering cross section has two distinct
terms: a spin-independent, or {\it scalar}, term, and 
a spin-dependent, or {\it axial} term.  It has been well
established that for nuclei with $A > 30$--50 ($A \equiv$ 
number of nucleons), the scalar
piece of the interaction tends to dominate the \chiN
scattering rate; however, there are significant regions of parameter
space where this is not so and the axial rate dominates
\cite{r:jkg,r:bednayakov}.  The importance of understanding the
axial \chiN interaction is amply demonstrated by a recent
SUSY interpretation of a Fermilab scattering event \cite{r:kane}.
These papers claim that the \neut might be an almost pure 
higgsino whose couplings to ordinary matter are completely
dominated by the axial part.  In this paper, we shall deal
with the axial \chiN interaction.  The relevant
nuclear physics for the scalar interaction is well approximated
by a fairly simple form factor, suitable for all nuclei\cite{r:epv,r:qtda}.
The axial response is far more complicated and requires detailed 
nuclear models.

\section{Nuclear Physics Input}

A variety of nuclear models have been used to calculate the axial
response of nuclei used as targets in dark matter detectors.
The conventional nuclear shell model \cite{r:bandw} has proven 
highly successful at accurately representing this response
when a reasonable nuclear Hamiltonian is used in a sufficiently
large model space \cite{r:mege73,r:meal27,r:ps,r:eopv}.  Until
recently, both of these ingredients have been absent for nuclei
in the $3s2d1g_{7/2}1h_{11/2}$ shell, including most of 
those included in this study.
With recent advances in computer power and storage, we can now construct
model spaces that contain most of the nuclear configurations that are
likely to dominate the spin response of nuclei such as \i127.
Coupled with this ability to perform sufficiently large 
calculations is the recent development of several realistic
nucleon-nucleon ($nn$) potentials \cite{r:hko,r:nim2}.  These
potentials can then be converted into suitable nuclear
interaction Hamiltonians via the G-matrix/Folded Diagram
technique \cite{r:hko}.  In this paper we consider two such nuclear
interactions, one using the Bonn A \cite{r:hko}
and the other the Nijmegen II \cite{r:nim2} $nn$ potential. 
The diagonalization of the Hamiltonian was performed using the
shell model code ANTOINE \cite{r:antoine}.

\subsection{The Hamiltonian}

The residual nuclear interaction based upon the Bonn A potential has been
described in Ref. \cite{r:hko}.  This Hamiltonian has
been derived for the model space consisting of the
$1g_{7/2}, \, 2d_{5/2}, \, 3s_{1/2}, \, 2d_{3/2},$ and
$1h_{11/2}$ orbitals (which we use in this study).  It was originally
derived to describe light Sn isotopes ($A \approx 102$--110) which
have no protons in the space.  In order to find good agreement
with observables for nuclei with $A \approx 130$, the single
particle energies (SPEs) were adjusted.  We made an initial guess at
the SPEs based upon the excited state energy
spectra of nuclei with either a single neutron hole in the
space ($^{131}$Sn) or a single proton in the space
($^{133}$Sb and $^{125}$Sb).  These initial SPEs were then
used in conjunction with the Two Body Matrix Elements (TBMEs)
of the interaction to calculate observables for 
the nucleus \i127.  We varied the SPEs
until reasonable agreement between calculation
and experiment was found for the following \i127  observables:
the magnetic moment ($\mu$), the low lying excited state energy
spectrum, and the quadrupole moment ($Q_{20}$).  This procedure
is similar to that performed in Ref. \cite{r:mege73}.
The magnetic moment
is extremely important, as it is the observable most closely
related to the \chiN scattering matrix element and has traditionally
been used as a benchmark of a calculation's accuracy.
In Fig.~\ref{f:fig1}, we show the final SPEs 
used in our calculations. 
In Table~\ref{t:tab1}
we show the final calculated values of $\mu$ and $Q_{20}$ for \i127
vs. the experimental values; agreement is excellent.  Once
the SPEs are specified, we have a reasonable Hamiltonian
to use for the nuclei we are studying.

In order to test the sensitivity of our results to the nuclear
Hamiltonian, we have also examined another one, derived from
the Nijmegen II $nn$ potential \cite{r:nim2}.  We have used the
codes and methods described in \cite{r:dcz} to convert the 
potential to a usable shell model interaction.  The procedure
is similar to that used for the Bonn A force.  The two sets
of TBMEs are generally similar but significant differences
do exist.  We initially used the same set of SPEs as above
but found that a significant lowering of the $1g_{7/2}$
SPE was necessary in order to find agreement with the observables.
The SPEs and comparisons with observables for this force are shown
in Fig. \ref{f:fig1} and Tables~\ref{t:tab1} and \ref{t:tab2}.

\subsection{The Model Space}

To perform a full basis, positive parity, calculation of the
\i127 ground state properties in the space consisting of the
$1g_{7/2}, \, 2d_{5/2}, \, 3s_{1/2}, \, 2d_{3/2},$ and
$1h_{11/2}$ orbitals, we would need to have basis states consisting
of roughly $1.3 \times 10^9$ Slater Determinants (SDs).
Current, state of the art, calculations  (including those presented here)
can diagonalize matrices with basis dimensions in the range
1--2$\times 10^7$ SDs; clearly severe truncations of the model
space are needed.  Fortunately, given the size of the model
spaces that can be treated, a truncation scheme that includes
the majority of relevant configurations can be devised.  

Our scheme is best
understood by viewing Fig.~\ref{f:fig1}.  As a base configuration, we have
for protons: $(1g_{7/2}2d_{5/2})^3$ (i.e. a total of 3 protons spread
among the $1g_{7/2}$ and $2d_{5/2}$ orbitals) and for neutrons: 
$(1g_{7/2}2d_{5/2})^{14} + (3s_{1/2}2d_{3/2})^{6} +
(1h_{11/2})^4$.  We then assign the following values of the
``jump'' to each orbital: jump($1g_{7/2},2d_{5/2}$) = 0,
jump($3s_{1/2},2d_{3/2}$) = 1, and jump($1h_{11/2}$) = 2.
The differences in these values is the cost of moving particles
between the different (sets of) orbitals.
Hence, to move a proton from the $1g_{7/2}$ to the $2d_{5/2}$
costs nothing while moving one from the $1g_{7/2}$ to the $3s_{1/2}$
would cost 1 unit of jump (to the $1h_{11/2}$ would cost 2 units).
It would cost 2 units of jump to move 2 neutrons from the 
$2d_{3/2}$ to the $1h_{11/2}$, etc....  All that remains is to
specify the total amount of jump available.  In our truncation,
we allow protons up to 3 units of jump, neutrons up to 4 units,
and a total of up to 4 when adding the jump used by the protons
plus neutrons.  Thus, if the protons remain in the $1g_{7/2}$ and $2d_{5/2}$
orbitals (as they tend to do), the following neutron
configurations are allowed: $(1g_{7/2}2d_{5/2})^{14} + 
(3s_{1/2}2d_{3/2})^{6} + (1h_{11/2})^4$, 
$(1g_{7/2}2d_{5/2})^{14} + (3s_{1/2}2d_{3/2})^{4} +
(1h_{11/2})^6$, 
$(1g_{7/2}2d_{5/2})^{13} + (3s_{1/2}2d_{3/2})^{5} +
(1h_{11/2})^6$, 
$(1g_{7/2}2d_{5/2})^{14} + (3s_{1/2}2d_{3/2})^{2} +
(1h_{11/2})^8$, 
and $(1g_{7/2}2d_{5/2})^{12} + (3s_{1/2}2d_{3/2})^{6} +
(1h_{11/2})^6$.  If 1 or 2 protons are excited out of the 
$1g_{7/2}$ and $2d_{5/2}$ orbitals, the last 2 neutron configurations
are not allowed.  In this truncation, the m-scheme dimension of
the \i127 model space is about 3 million SDs.

Our results indicate that this space is more than adequate to describe
the ground state properties of the nuclei considered.  As mentioned
above, our calculation of the observables agrees well with experiment.
The major potential problem with this model space would be
if it failed to allow enough neutron excitations into the $1h_{11/2}$ orbital
.  It allows at most 8 neutrons out of a possible 12 in that
orbital.  In Table~\ref{t:tab3}  we present the occupation numbers for
\i127.  We see that our interactions do not seem to prefer
excitation of more than one extra neutron pair to 
the $1h_{11/2}$.  Most configurations
have six neutrons in that orbital, while eight are 
allowed.  Hence, our model space is 
more than adequate.

For the two Xenon isotopes considered ($A = 129$ and 131), we have used
exactly this truncation scheme.  For \t125 we used this scheme
and also one where the total jump and total neutron jump was 6 (instead
of 4).  Very little difference was noticed for the two truncations.
In this paper we present the results for the larger truncation since it
should be slightly more realistic.

\section{Results}

\subsection{The Zero Momentum Transfer Limit}

Neutralinos in the halo of our Galaxy are characterized by
a mean virial velocity of,
$v \simeq \langle v \rangle  \simeq 300\,{\rm km/sec} =
10^{-3} c$.  The maximum characteristic momentum transfer in \chiN
scattering is $q_{max} = 2 M_r v$  where $M_r$ is
the reduced mass of the \chiN  system.  If the product
$q_{max}R$ is small ($\ll 1$), where $R$ is the nuclear size,
the matrix element for spin-dependent \chiN scattering reduces to a very
simple form \cite{r:jkg,r:epv}
\begin{equation}
 {\cal M} = C \langle N\vert a_p {\bf S}_p + a_n {\bf S}_n
\vert N \rangle \cdot {\bf s}_{\tilde \chi} 
\end{equation}
where
\begin{equation}
 {\bf S}_i = \sum_k {\bf s}_i ({\bf k}), \;\;\;\; i = p,n
\end{equation}
is the total nuclear spin operator, $k$ is a sum over all
nucleons, and $a_p,\, a_n$ are
\2nu-nucleon coupling constants which depend
upon the quark spin-distribution within the nucleons and on the
composition of the \2nu.
Much of the uncertainties arising
from electroweak and QCD scale physics are encompassed by
$a_p$ and $a_n$.  The normalization $C$ involves the coupling
constants, masses of the exchanged bosons and various LSP
mixing parameters that have no effect upon the nuclear matrix
element.  
Eq. (2) has often been written as
\begin{equation}
 {\cal M} = C \Lambda \langle N\vert {\bf J}
\vert N \rangle \cdot {\bf s}_{\tilde \chi} 
\end{equation}
with
\begin{equation}
 \Lambda = {{\langle N\vert a_p {\bf S}_p + a_n {\bf S}_n
\vert N \rangle}\over{\langle N\vert {\bf J}
\vert N \rangle}} =
{{\langle N\vert ( a_p {\bf S}_p + a_n {\bf S}_n ) \cdot {\bf J}
\vert N \rangle}\over{ J(J+1)
}}. 
\end{equation}
Examples of the full \chiN cross section can be found in 
Refs.~\cite{r:jkg,r:epv,r:mege73}.

Equations (2-5) show that the \neut couples to the spin carried
by the protons and the neutrons.  The matrix element (2) is similar
to the magnetic moment operator:
\begin{equation}
\mu =  \langle N\vert g_n^s {\bf S}_n + g_n^l {\bf L}_n +
g_p^s {\bf S}_p + g_p^l {\bf L}_p\vert N \rangle.
\end{equation}
The {\it free particle} $g$-factors are given by:
$g_n^s = -3.826, \, g_n^l = 0, \, g_p^s = 5.586$, and
$g_p^l = 1$ (in nuclear magnetons).
Given the similarities of eqs. (2) and (6), it is no surprise 
that  $\mu$ is often used as a benchmark on the accuracy of 
the calculation of \sp and \sn in $\Lambda$.  We follow
that prescription as well.  In the following section we will briefly
outline some problems with this procedure.

In Table~\ref{t:tab2} we present the values for:
\sp, \sn, ${\bf L}_p$, ${\bf L}_n$, and $\mu$ that we calculate
for each Hamiltonian for the nuclei \i127, \x129, \131x, and
\t125.  We also include the experimentally measured
magnetic moment, it is apparent that agreement is quite good
for all nuclei.  A number of other calculations
of these quantities appear in the literature, and we include
summaries of these calculations in the table as well.
The following abbreviations are used for the 
various nuclear models:
Bonn A $\equiv$ our calculation using the Bonn A derived force,
Nijmegen II $\equiv$ our calculation using the Nijmegen II derived force,
OGM $\equiv$ the Odd Group Model\cite{r:ogm},
IBFM $\equiv$ the Interacting Boson Fermion Model \cite{r:ibfm},
ISPSM $\equiv$ the Independent Single Particle Shell Model \cite{r:ellis},
TFFS $\equiv$ the Theory of Finite Fermi Systems \cite{r:tffs}, and
QTDA $\equiv$ the Quasi Tamm-Dancoff Approximation \cite{r:qtda}.  
In most previous experimental analyses, the OGM values have been
used~\cite{r:italy,r:boulby}.

Examining our results for \sp and \sn in Table~\ref{t:tab2} 
and comparing them
to results from other nuclear models reveal several interesting facts.
In almost every instance, our results show that the spin
$\vert {\bf S}_i \vert $ ($i = p,n$) carried by the unpaired
nucleon is greater than that found in the other nuclear
models (except for the ISPSM, where $\vert {\bf S}_i \vert $
is maximal).  Despite our larger values for $\vert {\bf S}_i \vert $,
our calculations have significant quenching of the magnetic moment
and are in good agreement with experiment in all cases (see
the later section on quenching).  The reason that we find larger
values of $\vert {\bf S}_i \vert $ for the odd group 
is due to the fact that we allow more excitation of the
even group of the nuclei; allowing them to be a major
contributor to the total nuclear spin:
${\bf J} =$ \sp + \sn + ${\bf L}_p$ + ${\bf L}_n$.  
The naive expectation for ${\bf L}_n$ in the Bonn A calculation
of \i127  is zero.  We find
${\bf L}_n = 0.779$, ${\bf L}_n$ is responsible for over 30\% of
iodine's total angular momentum (${\bf J} ={5\over{2}}$).  
This explains both the large
quenching of $\mu$ (${\bf L}_n$ does not contribute to $\mu$ 
since $g_n^l = 0$) and the large value of \sp found.
We note that most previous experimental analyses used
the OGM value for \i127, \sp = 0.07.  Our results give a factor 
of $\sim 20$ increase in iodine's sensitivity to spin-dependent
scattering over that previously assumed.  Due to the form
factor suppression (discussed below) a sodium
iodide detector's \cite{r:bottino,r:boulby} spin response is still
dominated by \n23 but not to the extent previously thought.
For the remainder of the nuclei considered Table~\ref{t:tab2} also reveals
increased scattering sensitivity, although the factor of increase
is much more modest.

\subsection{Quenching and Uncertainties}

As we noted earlier, the comparison of the computed
magnetic moment vs. the experimental value has been
used as the primary (and in some cases, only) indicator of
a calculation's reliability.  This seems quite reasonable in
light of the similarities between the matrix elements in eqs.
(2) and (6).  This prescription is not without several potential
problems \cite{r:mege73,r:engelshef}.  Not only does $\mu$ depend
upon the orbital angular momentum ${\bf L}_i$ but the spin
angular momentum  ${\bf S}_i$ is subtly different.
The \chiN matrix element (6) results from the non-relativistic
reduction of the axial-vector current.  Because of this, it is
not strongly affected by meson exchange currents (MECs).  The
magnetic moment's spin operators, ${\bf S}_i$, are a result of the
non-relativistic reduction of the vector current.  They
can be strongly affected by MECs \cite{r:meal27,r:engelshef}.
The effects of MECs upon $\mu$ is typically lumped together with
several other effects to give effective $g$-factors
\cite{r:sdmu,r:fpmu}.  Unfortunately, there is no hard and
fast rule as to what effective $g$-factors are the best.  
We have chosen to remain with the free particle
$g$-factors.  As an example of the potential uncertainties this
ambiguity leads to, we have also included, in 
Table~\ref{t:tab2}, the calculated
magnetic moment for our nuclei using a reasonable set of effective
$g$-factors.  The ``quenched'' magnetic moments are the values
in parentheses in the table and the effective $g$-factors used
are:  $g_n^s = -2.87, \, g_n^l = -0.1, \, g_p^s = 4.18$, and
$g_p^l = 1.1$.  The table shows that these $g$-factors do little,
overall, to improve the concordance between calculation and
experiment.

A related concern involves the quenching of the (isovector) Gamow-Teller (GT)
$g$-factor, $g_A$ \cite{r:mege73,r:engelshef}.  The spin term
of the GT operator also comes from the axial vector current and
thus is closely related to the spin operators in eq. (6).
Its is well established that most nuclear model calculations of
GT strength require a reduction of $g_A$ of order 20\%
\cite{r:bandw}.  Whether
this quenching of $g_A$ should also be applied to $a_1$ (the isovector
\2nu-nucleon coupling constant) is unknown
\cite{r:engelshef}.  Since there is no real guidance, and our magnetic
moments agree well with experiment, we do not believe that any
extra quenching of the spin matrix elements (or equivalently the
coupling constants $a_0$ and $a_1$) is desirable for these nuclei
when calculating \chiN scattering rates.  Nonetheless, as pointed
out in Ref.~\cite{r:engelshef} it is useful to keep these potential
uncertainties in mind when calculating scattering rates.

\subsection{Finite Momentum Transfer}

When the LSP was first proposed as a viable dark matter
candidate, its preferred mass was between 5 and 10 GeV \cite{r:dm1st}.
With a mass of this order and a typical galactic halo
velocity ($v \simeq 10^{-3} c$), the neutralino's total
momentum ($q \sim M_r v \sim 10 \,{\rm MeV}$) was small compared
to the inverse of the nuclear size ($1/R \sim 1/1\,{\rm fm} \sim
200\, {\rm MeV}$) and the zero momentum transfer limit was appropriate
for studies of \chiN scattering.  Since then,
experiments at accelerators have pushed
the allowed \neut mass, $m_{\tilde \chi}$, to larger values (there
are ways around this if some of the theoretical assumptions are
relaxed \cite{r:gandr}),  and it has been shown that
heavy \2nu's are just as viable as a dark matter candidate as
the lighter ones\cite{r:gkw,r:olive}.  As $m_{\tilde \chi}$
becomes larger than a few 10's of GeV the product
$qR$ starts to become non-negligible and finite momentum transfer
must be considered for heavier nuclei.  

The formalism for elastic \chiN scattering at all momentum
transfers has been developed in Refs.~\cite{r:epv,r:qtda}.  
Here, we follow precisely the definitions used in \cite{r:mege73}.
It is a simple matter to go from our definitions to those used
in Ref.~\cite{r:jkg}.
The formalism is a
straight forward extension of that developed for the
study of weak and electromagnetic semi-leptonic interactions
in nuclei \cite{r:donnelly}.  The differential
\chiN cross section is given by
\begin{equation}
{{d \sigma}\over{d q^2}} = {{8 G_F^2}\over{(2J + 1) v^2}} S(q),
\end{equation}
where $S(q)$ is the spin structure function
\begin{equation}
S(q) = \sum_{L\, odd} \big( \vert\langle N \vert\vert {\cal T}^{el5}_L
(q) \vert\vert N \rangle\vert^2 +
\vert\langle N \vert\vert {\cal L}^5_L
(q) \vert\vert N \rangle\vert^2\big) . 
\end{equation}
${\cal T}^{el5}(q)$ and ${\cal L}^5(q)$ are the transverse
electric and longitudinal multipole projections of the
axial-vector current operator \cite{r:donnelly}.
The double vertical lines imply that these are the reduced
matrix elements of these operators.
For their explicit form in the \neut context, see 
\cite{r:epv,r:qtda,r:mege73}.
In the limit of zero momentum transfer $S(q)$ reduces to
\begin{equation}
S(0) = {2 J + 1\over{\pi}} \Lambda^2 J(J + 1). 
\end{equation}

The reduced matrix elements of the multipoles in eq.~(8) are
easily evaluated in the harmonic oscillator basis in the
nuclear shell model \cite{r:donnelly}.  With the exception
of the calculation of the $^{27}$Al structure function in
\cite{r:meal27} all calculations of $S(q)$ have used bases of these
harmonic oscillator wave functions.  In this paper, we have
used the more realistic Wood-Saxon wave functions to evaluate
eq.~(8).  To specify the wave functions, we use the parameters
recommended in \cite{r:bertsch}.  We have used the codes from
\cite{r:blok} to calculate the actual wave functions.  We
have also calculated the Bonn A structure function for \i127 using
harmonic oscillator wave functions.  The differences in the
two prescriptions are significant at very large momentum transfers
but are minor at most relevant values of of the momentum transfer ($q$).

It is useful (and traditional) to use the isospin convention
instead of the proton-neutron formalism when discussing 
\chiN scattering at finite momentum transfer.  Writing
the isoscalar coupling constant as $a_0 = a_n + a_p$
and the corresponding isovector coupling constant as
$a_1 = a_p - a_n$ we may split $S(q)$ into a pure
isoscalar term, $S_{00}$, a pure isovector term,
$S_{11}$, and an interference term, $S_{01}$, in the following
way:
\begin{equation}
S(q) = a_0^2 S_{00}(q) + a_1^2 S_{11}(q) + a_0 a_1 S_{01}(q).
\end{equation}
Using this decomposition of $S(q)$ it is a simple matter
to derive the structure function for a \neut of arbitrary
composition.

Two factors contribute to the maximum
allowed momentum transfer.  As $m_{\tilde \chi}$ becomes much greater
than the nuclear mass, $m_N$, the reduced mass asymptotes
to $M_r \rightarrow m_N$.  Also, the \2nu's have a
Maxwellian velocity distribution in the halo and some will
possess velocities significantly greater
than $\langle v \rangle \simeq 10^{-3}c$.
A maximum velocity
of $v_{max} \simeq 700$ km/sec (slightly greater than
Galactic escape velocity \cite{r:bernabei})  implies maximum
momentum transfers of $q_{max}(A \sim 127) \simeq 550$ MeV. 
This value is not {\it small}
compared to the inverse nuclear size.  In a harmonic
oscillator basis, the fiducial nuclear size is set by the
oscillator parameter,
$b = 1\,{\rm fm}\, A^{1/6} = (1/197.327 {\rm MeV}) A^{1/6}$.
In order to maintain contact with previous literature
\cite{r:mege73,r:meal27} we retain $b$ as the size parameter
in our Wood-Saxon evaluations of $S(q)$.  We do, however, use
a slightly better, empirical, parameterization of $b$ \cite{r:wbmb}:
$b = (41.467/\hbar\omega)^{1/2}$ fm with $\hbar\omega = 
45 A^{-1/3} - 25 A^{-2/3}$ MeV.  Hence, we have values near 
$b(A = 127) = 2.282$ fm $ = 1/86.47$ MeV for the nuclei in this study.
We parameterize all of our structure functions in terms
of $y \equiv (qb/2)^2$.  For $y \ll 1$ the effects of finite momentum
transfers are small; for $y \ge 1$ the effects are quite noticeable.
For these nuclei $y_{max} = (q_{max}b/2)^2 \simeq 10 \gg 1$, hence
nuclear form factors are extremely significant.  These extremely
large values of $y$ are only valid for extremely massive \2nu's
moving near escape velocity.  A more realistic \2nu, with 
$m_{\tilde \chi} = 100$ GeV moving at $\langle v \rangle$ would have
$y_{max} \simeq 0.4$.

In order to cover all of the relevant \neut parameter space, we
have evaluated the structure functions all the way to $y = 10$
for the nuclei studied.  This presents a problem, in that
it has become standard to present structure functions as polynomials
in $y$ of order 6 or less. (A structure function of this sort can
easily be incorporated into the code {\it Neutdriver} of Ref.
\cite{r:jkg}.)  We could find no suitable fits of this form valid
out to values of $y = 10$.  We have addressed this in two ways.
In appendix B we present fits of the structure functions $S_{00}$,
$S_{01}$, and $S_{11}$ as
6th order polynomials in $y$.  These fits are only good for
values of $y < 1$.  They should NOT be used beyond this value
as they give meaningless results.
In order to accurately represent $S(q)$ at all relevant momentum
transfers, we have had to resort to a somewhat more complicated
functional form.  In a harmonic oscillator basis, the matrix elements
of the operators ${\cal T}^{el5}(q)$ and ${\cal L}^5(q)$ are 
precisely represented as polynomials in $y$ times a factor of
exp(-$y$). (The isovector Goldberger-Trieman term in ${\cal L}^5(q)$
complicates this slightly.)  Using this form as a guide we have
have fit the structure functions as 8th order polynomials in
$y$ times a factor or exp(-$2y$)  This form has proven adequate
to accurately  describe the structure functions for 
\i127 and \131x.  A slightly more complicated
form with a term added to mimic the effect of the Goldberger-Trieman
term was required for \x129 and \t125.  As an example, we present the fit
for the term $S_{00}(q)$ for the Bonn A calculation 
of \i127:

\begin{eqnarray}
S_{00}(y) & = & e^{-2y} (0.0983393 - 0.489096 y + 1.1402 y^2 - 1.47168 y^3 +
1.1717 y^4 \nonumber \\
 & & \,\,\,\, - 0.564574 y^5 + 0.158287 y^6 - 0.0238874 y^7 + 0.00154252 y^8) 
\end{eqnarray}
We relegate the remaining formulae to appendix C.  The various
fits can be acquired in a form suitable for inclusion in a Fortran
program by contacting one of the authors.  

In Fig.~\ref{f:fig2} panels A--D, we present the 
functions $S_{ij}(y)$ for the
nuclei \i127, \x129, \131x, and \t125.  The solid lines are for the
calculations using the Bonn A Hamiltonian and the dashed
lines are for the Nijmegen II based Hamiltonian.  In order to
make comparisons with other work easier, we restrict the results to 
values of $y \leq 2$ ($q^2 \leq \sim 60000$ MeV$^2$).  For illustration,
in Fig.~\ref{f:fig3} we show the full structure function of \i127 out to
$y = 10$.  In Fig.~\ref{f:fig4} panels A--D, we show the full structure
functions for a pure $\tilde B$ ($Z_1 = 1, \; Z_2 = Z_3 =
Z_4 = 0$) for each of the nuclei out to
a value of $y = 2$.  In these figures each function has been
normalized to the value $S(y = 0) = 1$ in order to highlight
the similarities and differences in the shapes
of the structure functions.  
In our definition of the $\tilde B$ we use the older, EMC, values
of the spin content of the proton.  This convention makes it
easier to compare our work to previous work on \131x 
\cite{r:qtda,r:tffs2}.  The precise values of $a_p$ and $a_n$
(or $a_0$ and $a_1$) can be found in \cite{r:mege73}; the
ratio is: $a_0/a_1 = 0.297$.

In Figs.~\ref{f:fig4} and \ref{f:fig5}
all of the structure functions have been
normalized to $S(y = 0) = 1$ to highlight their
similarities and differences.  In order to correctly gauge
 the true differences between the various $S(q)$, the different
normalizations must be taken into account.  This is easily
done by using eq. (9) and Table~\ref{t:tab2}.  As an example, consider
\i127 in panel A of Fig.~\ref{f:fig4} and 
in Fig.~\ref{f:fig5}.  To truly compare the
structure functions each of the lines needs to be multiplied
by a factor such that the ratio at $y = 0$ is given by
Bonn A (Wood-Saxon) : Nijmegen II (Wood-Saxon) : Bonn A (Harmonic
Oscillator) : Phenomenological (w/ OGM) : Single Particle 
(Harmonic Oscillator)
= 1 : 1.42 : 1 : 1/14.7 : 3.47.  Similar results can be recovered for
the other nuclei considered.

The line labeled Phenomenological above and in the figures requires
some explanation.  This is a shape for a general structure function
postulated and used in Ref.~\cite{r:boulby}.  It is apparent
from the figures that this approximation does a reasonable
job in reproducing $S(q)$ for $y \le 2$.  It is clearly inadequate
for larger values of the momentum transfer.  Below $y = 2$, its
shortcomings are also clear but any result derived using
this parameterization of $S(q)$ for \i127 should not be
far off.  The overall \i127 axial result of Ref.~\cite{r:boulby}
is another matter, as that paper normalizes to the OGM at
$y = 0$.  As we have shown, that model severely underestimates
$S(0)$.

While the parameterization of $S(q)$ in \cite{r:boulby} is
adequate for \i127 (although we now advocate the use of the
\i127 structure functions presented in the appendices), it is not
applicable for all nuclei.  The flattening observed in
$S(q)$ near $y = 1$ is the result of higher order multipoles
becoming important in eq. (8).  For \i127 the $L = 1,3,5$ multipoles
all contribute to $S(q)$.  For small $y \, (\ll 1)$, the structure 
function is dominated by the $L = 1$ multipole. For $y \ge 1$, all
three multipoles contribute and the higher order multipoles dominate.
For $J = {1\over{2}}$ nuclei, such as \x129 and \t125, only the
$L =1$ multipole can contribute.  Figs.~\ref{f:fig2} and \ref{f:fig4}
panels B and D clearly show
that there is no flattening of $S(q)$.  Hence, an approximate form
like that in \cite{r:boulby} is clearly inappropriate in these
cases.  In Fig.~\ref{f:fig6} we show the Bonn A derived structure functions
for a pure $\tilde B$ for all four nuclei.  It is obvious that they
can not all be fit by a single, simple, parameterization.
Figs.~\ref{f:fig4} and \ref{f:fig5} 
do show that the pure single particle form factor
also does an acceptable, but not compelling, job of representing the
structure functions at all momentum transfers if {\it correctly
normalized} at $y = 0$.  The correct single particle form factor can be
easily found by using the tables in the paper by Donnelly and Haxton
of Ref.~\cite{r:donnelly}.

Examining the structure functions for \t125 and \x129 in
Fig.~\ref{f:fig6} illustrates an interesting feature.  Both of these
nuclei are $J = {1\over{2}}$ nuclei with an unpaired neutron.
In the ISPSM both of these nuclei would be represented by a
neutron in the 3$s_{1/2}$ orbital and have virtually
identical properties.  Table~\ref{t:tab2}  shows that
the magnetic moments are quite similar but that the distribution
of the angular momentum in each nucleus is quite different.
This is most obvious in the orbital angular momentum
${\bf L}_i$ where the two distributions are quite different.
Fig.~\ref{f:fig6} reveals that while the structure functions have
definite similarities, there are significant differences
as well.  We point all of this out to highlight the fact that
seemingly very similar nuclei can have very different properties
when examined in detail.  If precise information on the
spin distribution of a nucleus is required, detailed calculations
must be performed.

It is also useful to consider differences in $S(q)$ that are
the result of different nuclear models.  $S(q)$ has been calculated
for \131x in the context of two other nuclear models,
the QTDA \cite{r:qtda} and the TFFS \cite{r:tffs2}, as well as
here.  In Fig.~\ref{f:fig7} we show $S(q)$ for a 
pure $\tilde B$ as a function of
$q^2$ for \131x.  This figure is meant to be a direct analog
of Fig. 2 in Ref.~\cite{r:qtda} and Fig. 3 of Ref.~\cite{r:tffs2}.
Examining the 3 figures yields some interesting conclusions.
All three calculations show significant quenching compared to
the single particle estimate.  The spin distribution between the
QTDA and TFFS is somewhat different while the full structure functions
are quite similar.  While the values for \sn differ very little
between our work and the QTDA, the difference in the values of
$S(0)$ is almost a factor of 2 between the two calculations.
Finally, it should be noted that both the QTDA and TFFS calculations
of $S(q)$ asymptote to the single particle structure function.  This
is not the case in our calculations, which are well below the
single particle estimate for all values of $q^2$.  This can also be seen
in Ref.~\cite{r:engelshef} where our values of
$S_{ij}(q)$ for the Bonn A calculation are compared to those of the
QTDA calculation.  In that comparison, it is apparent that
the shell model derived structure functions have a much steeper
fall off as a function of $q^2$.  

Finally, we mention the difference between the structure functions
derived using Wood-Saxon wave functions vs. those derived using
a harmonic oscillator basis. In Fig.~\ref{f:fig4} panel A and 
Fig.~\ref{f:fig5}  we show the
structure functions for \i127 using both sets of basis states.
Significant differences between
the two sets are apparent for extremely high momentum
transfers but in the range that is most relevant for dark matter
detection there is little difference.

In this section we have discussed the formalism of, and presented
our results for, the \chiN axial structure function for several
nuclei involved in dark matter detectors.  Accurate fits
which are suitable for use in calculating event rates in 
detectors are presented in the appendices.  Several interesting
features of the functions have been noted and it is apparent that no
single simple parameterization of $S(q)$ is suitable for all
nuclei.  Finally we have compared our results to other calculations
of \131x structure functions and noted several similarities and
differences that arise from different nuclear models.

\section{Discussion}

In this paper we have calculated the full axial response for
several heavy nuclei used in a number of direct dark matter
detection experiments.  With this set of structure functions,
there now exists accurate calculations of the axial
\chiN response to most, if not all, nuclei used as targets
in dark matter detectors.  We have used the largest model
spaces practical in conjunction with realistic nuclear Hamiltonians
to construct our wave functions.  Two different nuclear Hamiltonians
have been used in order to investigate the sensitivity of
our results to this particular input.

The differences in the response due to the two forces is clearly
visible in Table~\ref{t:tab2}  and Figs.~\ref{f:fig2}--\ref{f:fig6}.  
In all cases, reasonable
agreement between calculation and experiment for the magnetic
moment (using free particle $g$-factors) is achieved.
It is obvious from the
table that the differences between the two calculations
are non-trivial but that they are quite a bit smaller
than the differences coming from the use of alternate nuclear
models.  This shows that the interaction is not the primary
uncertainty in calculations of the \chiN nuclear response.

We have also attempted to examine the uncertainty due to the
nuclear model chosen.  A number of calculations of \131x's response
have been performed.  We find that our calculations are in 
reasonable agreement with other studies of the spin distribution
and finite momentum response but distinct differences do exist.  In the
case of \131x it is not immediately obvious which calculation is to
be preferred.  The calculations presented here contain
more excitations within the model space and use
more modern and realistic nuclear interactions than the others
in the literature.  By restricting excitations within this
model space, the calculations presented in \cite{r:qtda} included 
excitations out of the space that we worked within.  Both 
calculations reproduce the magnetic moment well, with the
QTDA calculation doing slightly better.  (We note that the
QTDA model, and a refined version of it, have been applied to
\i127 and was unable to reproduce the magnetic moment with
sufficient accuracy~\cite{r:engeli127}.)

Another improvement incorporated into these calculations of $S(q)$
is the use of Wood-Saxon wave functions to evaluate the multipole
operators in eq. (8).  The Wood-Saxon wave functions made a significant
difference at extremely high momentum transfers when compared to
the usual harmonic oscillator wave functions.  At the more
modest momentum transfers typical of ``average'' neutralinos,
the difference is found to be small.

Now that these structure functions are available, we hope
that they will be used by all experiments based upon the
materials that we have studied.  This will facilitate
comparisons between different groups.  To date, each experiment
has used different structure functions in their analyses.  A
first step in this direction has already been taken by one
group. The most recent analysis of the experiments based
in the Gran Sasso laboratory uses the \i127 and \x129 structure
functions presented here\cite{r:bottino}.  
We hope that other groups will follow
suit so that all future results can be compared on equal footing.

\acknowledgements

M.T.R. gratefully acknowledges support from the Weingart
Foundation and the U.S. National Science Foundation
under Grants PHY94-12818 and PHY94-20470.
Oak Ridge National Laboratory is managed by Lockheed Martin Energy
Research Corp. for the U.S. Department of Energy under contract number
DE-AC05-96OR22464.  D.J.D acknowledges an E.~P.~Wigner Fellowship from ORNL.
We acknowledge useful discussions with Petr Vogel and Dao-Chen Zheng.

\vfill\eject

\begin{table}
\caption[Table I.]{The calculated magnetic and quadrupole moments
of \i127 compared to experiment for calculations using both
effective interactions.  For the quadrupole moment, effective
charges of $e_p = 1.5 e$ and $e_n = 0.5 e$ have been used.
The magnetic moment calculations use the free particle
$g$-factors.  We also include the ISPSM estimates of the quantities
in order to illustrate the quenching obtained.}
\begin{tabular}{ccccc}
Observable & ISPSM & Bonn A & Nijmegen II & Experiment \\
\tableline
$\mu$ & 4.79 & 2.775 & 3.150 & 2.813 \\
 $Q_{20}$ & -0.654 & -0.577 & -0.577 & -0.789 \\
\end{tabular}
\label{t:tab1}
\end{table}

\begin{table}
\caption[Table II.]{A comparison of various calculations 
of the spin distribution
of $^{127}$I, \x129, \131x, and \t125.  Bonn A and Nijmegen II 
are the calculations presented
here.  OGM is the Odd Group Model of \cite{r:ogm}. IBFM is
the Interacting Boson Fermion Model of \cite{r:ibfm}. TFFS
is the Theory of Finite Fermi Systems calculation of \cite{r:tffs}.
QTDA is the Quasi Tamm-Dancoff Approximation of \cite{r:qtda}.
Here, as elsewhere in the paper, we take the matrix elements of 
the operators between the maximally stretched states, e.g.
$\langle {\bf S}_p\rangle \equiv \langle J,M_J = J \vert {\bf S}_p
\vert J, M_J = J \rangle$.  A blank entry means that the value
of that particular angular momentum component was not presented
in the reference. An entry of N/A in the magnetic moment column
implies that the experimental magnetic moment was used to find the values
of spin $\langle {\bf S}_p\rangle$ or $\langle {\bf S}_n\rangle$
shown. Calculations of the magnetic moment using effective g-factors as
described in the text are given in parenthesis.}
\begin{tabular}{cccccc}
 & $\langle {\bf S}_p\rangle$ & $\langle {\bf S}_n\rangle$ & $\langle
{\bf L}_p\rangle$ & $\langle {\bf L}_n\rangle$ & $\mu$ \\
\tableline
\i127 & & & & & \\
\tableline
Experiment & & & & & 2.813 \\
Bonn A & 0.309 & 0.075 & 1.338 & 0.779 & 2.775 (2.470)  \\
Nijmegen II & 0.354 & 0.064 & 1.418 & 0.664 & 3.150 (2.790) \\
OGM~\cite{r:ogm} & 0.07  & 0.0 & 2.43 & 0.0 & N/A \\
IBFM~\cite{r:ibfm} & 0.166 & 0.01 & & & N/A \\
TFFS~\cite{r:tffs} & 0.15 &  & & &  \\
ISPSM~\cite{r:ellis} & 0.5 & 0.0 & 2.0 & 0.0 & 4.793 \\
\tableline
\x129 & & & & & \\
\tableline
Experiment & & & & & -0.778 \\
Bonn A & 0.028 & 0.359 & 0.227 & -0.114 & -0.983 (-0.634)  \\
Nijmegen II & 0.0128 & 0.300 & 0.372 & -0.185 & -0.701 (-0.379) \\
OGM~\cite{r:ogm} & 0.0  & 0.2 & 0.0 & 0.3 & N/A \\
IBFM~\cite{r:ibfm} & 0.0 & 0.2 & & & N/A \\
TFFS~\cite{r:tffs} &  & 0.25 & & &  \\
ISPSM~\cite{r:ellis} & 0.0 & 0.5 & 0.0 & 0.0 & -1.913 \\
\tableline
\131x & & & & & \\
\tableline
Experiment & & & & & 0.69 \\
Bonn A & -0.009 & -0.227 & 0.165 & 1.572 & 0.980 (0.637)  \\
Nijmegen II & -0.012 & -0.217 & 0.215 & 1.514 & 0.979 (0.347) \\
QTDA~\cite{r:qtda} & -0.041 & -0.236 & 0.026 & 1.751 & 0.70 \\
OGM~\cite{r:ogm} & 0.0  & -0.18 & 0.0 & 1.68 & N/A \\
IBFM~\cite{r:ibfm} & 0.0 & -0.17 & & & N/A \\
TFFS~\cite{r:tffs} &  & -0.186 & & &  \\
ISPSM~\cite{r:ellis} & 0.0 & -0.3 & 0.0 & 1.8 & 1.71 \\
\tableline
\t125 & & & & & \\
\tableline
Experiment & & & & & -0.889 \\
Bonn A & 0.001 & 0.287 & 0.077 & 0.135 & -1.015 (-0.749)  \\
Nijmegen II & -0.0003 & 0.323 & 0.102 & 0.075 & -1.134 (-0.824) \\
OGM~\cite{r:ogm} & 0.0  & 0.23 & 0.0 & 0.27 & N/A \\
IBFM~\cite{r:ibfm} & -0.0004 & 0.23 & & & N/A \\
TFFS~\cite{r:tffs} &  & 0.22 & & &  \\
ISPSM~\cite{r:ellis} & 0.0 & 0.5 & 0.0 & 0.0 & -1.913 \\
\end{tabular}
\label{t:tab2}
\end{table}

\begin{table}
\caption[Table III.]{The proton ($p$) and neutron ($n$) occupation 
numbers obtained for each orbital in the \i127 calculations.
The points to notice are the similarities between the two different
interactions and the fact that the $1h_{11/2} n$ number is
significantly less than 8, the maximum number allowed.}
\begin{tabular}{ccccc}
Orbital & Bonn A & Nijmegen II \\
\tableline
$1g_{7/2} p$ & 1.97979 & 1.83023 \\
$1g_{7/2} n$ & 7.87440 & 7.94902 \\
$2d_{5/2} p$ & 0.89648 & 0.95545 \\
$2d_{5/2} n$ & 5.92205 & 5.93768 \\
$3s_{1/2} p$ & 0.02859 & 0.09023 \\
$3s_{1/2} n$ & 1.57985 & 1.71861 \\
$2d_{3/2} p$ & 0.09511 & 0.12374 \\
$2d_{3/2} n$ & 2.62799 & 2.71644 \\
$1h_{11/2} p$ & 0.00004 & 0.00034 \\
$1h_{11/2} n$ & 5.99571 & 5.67825 \\
\end{tabular}
\label{t:tab3}
\end{table}

\vfill\eject

\appendix
\section{\n23}

All of the current dark matter detectors which use iodine as 
a target also use sodium.  The detectors are large sodium iodide
(NaI) crystals \cite{r:bernabei}.  Since a detailed calculation of
the axial response of \n23 has not appeared in the literature, we
present one here.  The nucleus \n23 lies in the middle of
the $sd$ shell and therefore is amenable to the same methods
applied to other $sd$-shell nuclei.  For our calculation
we perform the exactly analogous calculation to those done for
$^{29}$Si in Ref.~\cite{r:mege73} and $^{27}$Al in \cite{r:meal27};
including the use of harmonic oscillator wave functions.
The details of the calculation can be found in the above
references.

For \n23 we use an oscillator parameter of
$b = 1.6864\,{\rm fm} = (1/117.01)$MeV$^{-1}$.
For our adopted  maximum halo velocity of $v = 700$ km/sec we
have $y_{max} = 0.1875$.  A breakdown of the angular momentum
along with a comparison of the measured and calculated magnetic
moments is presented in Table~\ref{t:tabA1}; agreement is excellent.  
Table~\ref{t:tabA1}
also shows a significant difference in \sp from that predicted
in the OGM.  Finally, in the following equations we present fits
to the structure functions $S_{ij}(q)$ as 3rd order polynomials
in $y$ which are highly accurate to values well past $y_{max}$.

\begin{table}
\caption[Table A1.]{The decomposition of the angular momentum for
\n23 along with the calculated and experimental magnetic moments.}
\begin{tabular}{cccccc}
 & $\langle {\bf S}_p\rangle$ & $\langle {\bf S}_n\rangle$ & $\langle
{\bf L}_p\rangle$ & $\langle {\bf L}_n\rangle$ & $\mu$ \\
\tableline
\n23 & & & & & \\
\tableline
Experiment & & & & & 2.218 \\
Calculation & 0.2477 & 0.0198 & 0.9117 & 0.3206 & 2.2196 \\
OGM & 0.1566 & 0.0 & 1.3434 & 0.0 & N/A \\
\end{tabular}
\label{t:tabA1}
\end{table}

\begin{eqnarray}
S_{00}(y) & = & 0.0379935 - 0.174341 y + 0.378299 y^2 - 0.342962 y^3 \\
S_{01}(y) & = & 0.0646525 - 0.350289 y + 0.910031 y^2 - 0.985833 y^3 \\
S_{11}(y) & = & 0.0275013 - 0.169641 y + 0.507868 y^2 - 0.617985 y^3
\end{eqnarray}

\section{The Abbreviated Structure Functions}

The fits to $S(q)$ in this appendix are only valid for $y \leq 1$.
The fits are presented as tables of the coefficients of
6th order polynomials in $y$: $S_{ij}(q) = \sum_{k = 0}^6 
C_k y^k$.  The first column gives the order of $y^k$ then next
3 columns give the corresponding values of the $C_k$ for $S_{00}$,
$S_{01}$, and $S_{11}$ for the Bonn A calculation.  The last
3 columns present the results for the Nijmegen II calculation
in the same manner.

\begin{table}
\caption[Table B1.]{\i127}
\begin{tabular}{ccccccc}
 & & Bonn A & & & Nijmegen II & \\
\tableline
 & $S_{00}$ & $S_{01}$ &
$S_{11}$  & $S_{00}$ & $S_{01}$ & $S_{11}$ \\
\tableline
$1$ & 0.0982724 & 0.119851 & 0.0365375 & 0.116548 & 0.161931 & 0.0562404 \\
$y$ & -0.675013 & -0.843567 & -0.262676 & -0.792274 & -1.14026 & -0.408512 \\
$y^2$ & 2.13531 & 2.73535 & 0.875115 & 2.49846 & 3.71441 & 1.37775 \\
$y^3$ & -3.7595 & -4.93029 & -1.61455 & -4.38312 & -6.71583 & -2.57019 \\
$y^4$ & 3.77735 & 5.05806 & 1.69076 & 4.38495 & 6.89384 & 2.70866 \\
$y^5$ & -2.0091 & -2.73609 & -0.930164 & -2.32223 & -3.72586 & -1.4945 \\
$y^6$ & 0.435566 & 0.60084 &  0.206944 & 0.501504 & 0.817068 & 0.332885 \\
\end{tabular}
\label{t:tabB1}
\end{table}

\begin{table}
\caption[Table B2.]{\x129}
\begin{tabular}{ccccccc}
 & & Bonn A & & & Nijmegen II & \\
\tableline
 & $S_{00}$ & $S_{01}$ &
$S_{11}$  & $S_{00}$ & $S_{01}$ & $S_{11}$ \\
\tableline
$1$ & 0.0712796 & -0.121583 & 0.0518388 & 0.0464592 & -0.0853234 & 0.0391694 \\
$y$  & -0.480418 & 0.874546 & -0.394855 & -0.313776 & 0.614961 & -0.299123 \\
$y^2$ & 1.47263 & -2.83165 & 1.34334 & 0.965631 & -1.98471 & 1.00873 \\
$y^3$ & -2.53226 & 5.09221 & -2.51522 & -1.6666 & 3.54959 & -1.86483 \\
$y^4$ & 2.49681 & -5.19757 & 2.64796 & 1.64774 & -3.60225 & 1.93996 \\
$y^5$ & -1.30712 & 2.79235 & -1.4557 & -0.864196 & 1.92573 & -1.05644 \\
$y^6$ & 0.279589 & -0.60881 & 0.322793 & 0.185069 & -0.418234 & 0.232626 \\
\end{tabular}
\label{t:tabB2}
\end{table}

\begin{table}
\caption[Table B3.]{\131x}
\begin{tabular}{ccccccc}
 & & Bonn A & & & Nijmegen II & \\
\tableline
 & $S_{00}$ & $S_{01}$ &
$S_{11}$  & $S_{00}$ & $S_{01}$ & $S_{11}$ \\
\tableline
$1$   & 0.0295866 & -0.0544505 & 0.0250499 & 0.0277038 & -0.0497326 & 
0.0223178 \\
$y$   & -0.185155 & 0.36762 & -0.181162 & -0.175382 & 0.338942 & -0.162659 \\
$y^2$ & 0.593387 & -1.18133 & 0.593168 & 0.560377 & -1.10015 & 0.542687 \\
$y^3$ & -1.03518 & 2.05291 & -1.03886 & -0.996936 & 1.97087 & -0.98921 \\
$y^4$ & 1.00492 & -1.98269 & 1.00706 & 1.01 & -1.99963 & 1.01495 \\
$y^5$ & -0.507773 & 0.996715 & -0.50709 & -0.540224 & 1.06809 & -0.54588 \\
$y^6$ & 0.103658 & -0.202596 & 0.103134 & 0.11739 & -0.231591 & 0.118858 \\
\end{tabular}
\label{t:tabB3}
\end{table}

\begin{table}
\caption[Table B4.]{\t125}
\begin{tabular}{ccccccc}
 & & Bonn A & & & Nijmegen II & \\
\tableline
 & $S_{00}$ & $S_{01}$ &
$S_{11}$  & $S_{00}$ & $S_{01}$ & $S_{11}$ \\
\tableline
$1$   & 0.0396831 & -0.0788638 & 0.0391772 & 0.049567 & -0.0993273 & 
0.0497519 \\
$y$   & -0.271174 & 0.572717 & -0.30043 & -0.342464 & 0.731525 & -0.387508 \\
$y^2$ & 0.869383 & -1.90069 & 1.03775 & 1.06657 & -2.39301 & 1.32901 \\
$y^3$ & -1.56951 & 3.46977 & -1.94604 & -1.85469 & 4.32285 & -2.49519 \\
$y^4$ & 1.61835 & -3.5546 & 2.02635 & 1.84644 & -4.42823 & 2.6334 \\
$y^5$ & -0.879731 & 1.90199 & -1.09438 & -0.976196 & 2.39046 & -1.45353 \\
$y^6$ & 0.194048 & -0.411614 & 0.238043 & 0.210989 & -0.524466 & 0.32414 \\
\end{tabular}
\label{t:tabB4}
\end{table}

\section{The Full Structure Functions}

The fits to $S(q)$ in this appendix are good for all values
of $y \le 10$.
The fits are presented as tables of the coefficients of
8th order polynomials in $y$ plus a term included to
mimic the Goldberger-Trieman term present in the longitudinal
multipole~\cite{r:qtda}, all multiplied by a factor
of $\exp(-2 y)$ : $S_{ij}(q) = (\sum_{k = 0}^8 
C_k y^k + C_9 {1\over{1 + y}}) e^{-2y}$.  The first column gives the 
order of $y^k$ then next
3 columns give the corresponding values of the $C_k$ for $S_{00}$,
$S_{01}$, and $S_{11}$ for the Bonn A calculation.  The last
3 columns present the results for the Nijmegen II calculation
in the same manner.  An example of the table's use can be found
by comparing Eq. (11) to the entries for $S_{00}$ in the
Bonn A portion of the \i127 table.

\begin{table}
\caption[Table C1.]{\i127}
\begin{tabular}{ccccccc}
 & & Bonn A & & & Nijmegen II & \\
\tableline
$\times (e^{-2 y}) $ & $S_{00}$ & $S_{01}$ &
$S_{11}$  & $S_{00}$ & $S_{01}$ & $S_{11}$ \\
\tableline
$1$ & 0.0983393 & 0.11994 & 0.0365709 & 0.11663 & 0.162054 & 0.056287 \\ 
$y$ & -0.489096 & -0.618424 & -0.194994 & -0.572149 & -0.836288 & -0.303825 \\
$y^2$ & 1.1402 & 1.50893 & 0.504876 & 1.33797 & 2.05944 & 0.794783 \\
$y^3$ & -1.47168 & -2.07367 & -0.747451 & -1.72517 & -2.83193 & -1.17027 \\
$y^4$ & 1.1717 & 1.77307 & 0.704334 & 1.37742 & 2.39726 & 1.06373 \\
$y^5$ & -0.564574 & -0.903597 & -0.393018 & -0.669986 & -1.21214 & -0.571342 \\
$y^6$ & 0.158287 & 0.26002 & 0.121881 & 0.190522 & 0.348612 & 0.172197 \\
$y^7$ & -0.0238874 & -0.0387025 & -0.0191881 & -0.0291803 & -0.0521813 & 
-0.0266165 \\
$y^8$ & 0.00154252 & 0.00235675 & 0.00121021 & 0.0019081 & 0.00320731 & 
0.00166238 \\
${1\over{1 + y}}$ & 0.0 & 0.0 & 0.0 & 0.0 & 0.0 & 0.0 \\
\end{tabular}
\label{t:tabC1}
\end{table}

\begin{table}
\caption[Table C2.]{\x129}
\begin{tabular}{ccccccc}
 & & Bonn A & & & Nijmegen II & \\
\tableline
$\times (e^{-2 y}) $ & $S_{00}$ & $S_{01}$ &
$S_{11}$  & $S_{00}$ & $S_{01}$ & $S_{11}$ \\
\tableline
$1$   & 0.0713238 & -0.12166 & -2.05825 & 0.046489 & -0.0853786 & -1.28214 \\ 
$y$   & -0.344779 & 0.644351 & 1.80756 & -0.225507 & 0.453434 & 1.09276 \\
$y^2$ & 0.755895 & -1.52732 & -1.27746 & 0.499045 & -1.06546 & -0.712949 \\
$y^3$ & -0.933448 & 2.02061 & 0.654589 & -0.622439 & 1.3867 & 0.314894 \\
$y^4$ & 0.690061 & -1.57689 & -0.221971 & 0.46361 & -1.0594 & -0.0835104 \\
$y^5$ & -0.302476 & 0.723976 & 0.0454635 & -0.20375 & 0.47576 & 0.0105933 \\
$y^6$ & 0.0765282 & -0.190399 & -0.00425694 & 0.0510851 & -0.122077 & 
0.000233709 \\
$y^7$ & -0.0103169 & 0.0263823 & -0.000136779 & -0.00670516 & 0.0164292 & 
-0.000243292 \\
$y^8$ & 0.000573919 & -0.00148593 & 0.00004396 & 0.00035659 & -0.000894498 & 
0.0000221666 \\
${1\over{1 + y}}$ & 0.0 & 0.0 & 2.11016 & 0.0 & 0.0 & 1.32136 \\
\end{tabular}
\label{t:tabC2}
\end{table}

\begin{table}
\caption[Table C3.]{\131x}
\begin{tabular}{ccccccc}
 & & Bonn A & & & Nijmegen II & \\
\tableline
$\times (e^{-2 y}) $ & $S_{00}$ & $S_{01}$ &
$S_{11}$  & $S_{00}$ & $S_{01}$ & $S_{11}$ \\
\tableline
$1$   & 0.0296421 & -0.0545474 & 0.0250994 & 0.0277344 & -0.0497844 & 
0.0223447 \\ 
$y$   & -0.133427 & 0.271757 & -0.137716 & -0.124487 & 0.247247 & -0.122063 \\
$y^2$ & 0.377987 & -0.723023 & 0.366609 & 0.328287 & -0.632306 & 0.319493 \\
$y^3$ & -0.579614 & 1.0545 & -0.53851 & -0.481399 & 0.896416 & -0.466949 \\
$y^4$ & 0.578896 & -0.971333 & 0.492545 & 0.475646 & -0.816445 & 0.428767 \\
$y^5$ & -0.345562 & 0.538422 & -0.269903 & -0.285177 & 0.452352 & -0.236789 \\
$y^6$ & 0.115952 & -0.168988 & 0.0836943 & 0.0968193 & -0.142686 & 0.0740837 \\
$y^7$ & -0.0201178 & 0.027416 & -0.0133959 & -0.0170957 & 0.0233463 & 
-0.0119668 \\
$y^8$ & 0.00141793 & -0.00180527 & 0.000868668 & 0.00123738 & -0.00156293 & 
0.000787042 \\
${1\over{1 + y}}$ & 0.0 & 0.0 & 0.0 & 0.0 & 0.0 & 0.0 \\
\end{tabular}
\label{t:tabC3}
\end{table}

\begin{table}
\caption[Table C4.]{\t125}
\begin{tabular}{ccccccc}
 & & Bonn A & & & Nijmegen II & \\
\tableline
$\times (e^{-2 y}) $ & $S_{00}$ & $S_{01}$ &
$S_{11}$  & $S_{00}$ & $S_{01}$ & $S_{11}$ \\
\tableline
$1$   & 0.0397091 & -0.0789431 & 0.0392236 & 0.0495946 & -0.0993873 & 
-1.92941 \\ 
$y$   & -0.196101 & 0.42738 & -0.229376 & -0.247766 & 0.54303 & 1.68075 \\
$y^2$ & 0.472653 & -1.09331 & 0.622146 & 0.547656 & -1.28816 & -1.16336 \\
$y^3$ & -0.650229 & 1.55324 & -0.922531 & -0.665532 & 1.67206 & 0.586501 \\
$y^4$ & 0.541926 & -1.28933 & 0.784648 & 0.474621 & -1.26883 & -0.207302 \\
$y^5$ & -0.264563 & 0.618441 & -0.382445 & -0.199442 & 0.56728 & 0.0514094 \\
$y^6$ & 0.074891 & -0.16964 & 0.105709 & 0.0481866 & -0.145438 & -0.00869728 \\
$y^7$ & -0.0114632 & 0.0248165 & -0.0154157 & -0.00616326 & 0.0195887 & 
0.000870366 \\
$y^8$ & 0.000749022 & -0.00152108 & 0.000928651 & 0.000322728 & -0.00106519 & 
0.0000354095 \\
${1\over{1 + y}}$ & 0.0 & 0.0 & 0.0 & 0.0 & 0.0 & 1.97923 \\
\end{tabular}
\label{t:tabC4}
\end{table}

\begin{figure}
\caption[Figure 1] {A visual description of the \i127 model space.
See the text for specific details of the construction of the space.
The other nuclei studied use the same jump assignments and SPEs.
The left $1g_{7/2}$ SPE is that used for the Bonn A interaction and the one
on the right is used by the Nijmegen II interaction.
}
\label{f:fig1}
\end{figure}

\begin{figure}
\caption[Figure 2] {The three spin structure functions $S_{ij}(q)$ where
$i,j = 0,1$ for the 4 nuclei considered. The results using both effective
interactions are plotted.  Accurate fits to these structure functions
can be found in Appendices B and C. A) $S_{ij}(q)$ for \i127; the ordering
is: $S_{01} > S_{00} > S_{11}$ for each force.  
B) $S_{ij}(q)$ for \x129; the ordering
is: $S_{00} > S_{11} > S_{01}$ for each interaction.
C) $S_{ij}(q)$ for \131x; the ordering
is: $S_{00} > S_{11} > S_{01}$ for each interaction.
D) $S_{ij}(q)$ for \t125; the ordering
is: $S_{00} > S_{11} > S_{01}$ for each interaction.
}
\label{f:fig2}
\end{figure}

\begin{figure}
\caption[Figure 3] {Another view of panel A in Fig.~\ref{f:fig2}.
Here we have extended $S_{ij}(q)$ out to $y = 10$ and chopped off much
of the initial fall off from $S_{ij}(0)$ in order to highlight the
similarities and differences between the two sets of structure
functions. 
}
\label{f:fig3}
\end{figure}

\begin{figure}
\caption[Figure 4] {The spin structure function $S(q)$ for a pure
$\tilde B$ ($a_0/a_1 = 0.297$) for the 4 nuclei considered.
Wood-Saxon (W.S.) wave functions have been used.
The results using both effective interactions are plotted.
Additionally, the pure single particle estimate of $S(q)$ with
harmonic oscillator (H.O.) wave functions is included for comparison.
All structure functions have been normalized to $S(0) = 1$ in order
to better compare their intrinsic shapes.  To truly compare the
differences the functions need to be normalized using Eq. (9) and
the values in Table~\ref{t:tab2}.  A) $S(q)$ for \i127.  Also included
for comparison is the $S(q)$ used in \cite{r:boulby} and the results
for the Bonn A interaction using H.O. wave functions.
B) $S(q)$ for \x129.
C) $S(q)$ for \131x.
C) $S(q)$ for \t125.
}
\label{f:fig4}
\end{figure}

\begin{figure}
\caption[Figure 5] {Another view of panel A in Fig.~\ref{f:fig4}.
Here we have extended $S(q)$ out to $y = 10$ and chopped off much
of the initial fall off from $S(0)$ in order to highlight the
similarities and differences between the various structure
functions.
}
\label{f:fig5}
\end{figure}

\begin{figure}
\caption[Figure 6] {The Bonn A calculations of $S(q)$ for a pure $\tilde B$
for all 4 nuclei compared.  Note the very large differences near
$y = 1$ between the nuclei with $J = {1\over{2}}$ (\x129 and \t125) and
those with larger $J$.
}
\label{f:fig6}
\end{figure}

\begin{figure}
\caption[Figure 7] {The \131x structure function for a pure
$\tilde B$.  The single particle structure function has been
normalized to $S(0) = 1$.  The Bonn A and Nijmegen II calculations
have been correctly normalized relative to the single particle
model.  This figure is a direct analog of, and should be compared
to, Fig. 2 of Ref.~\cite{r:qtda} and Fig. 3 of Ref.~\cite{r:tffs2}.
The major differences between these calculations and those of
\cite{r:qtda} and \cite{r:tffs2} are: We find 
$S(0) \simeq 0.4$--$0.5$ vs. their values of $S(0) \simeq 0.25$
and both of the other model's structure functions asymptote to
the single particle model for $q^2 > 0.02$ GeV$^2$ while these
calculations stay well below the single particle model.
}
\label{f:fig7}
\end{figure}

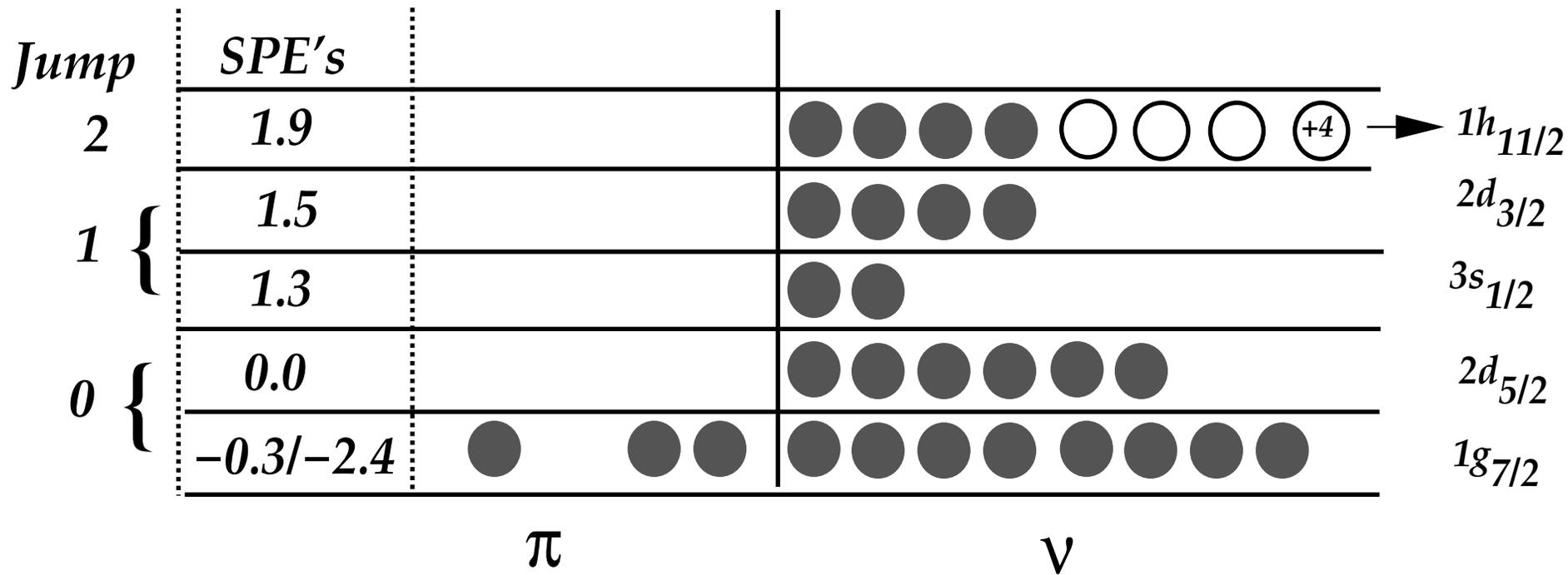
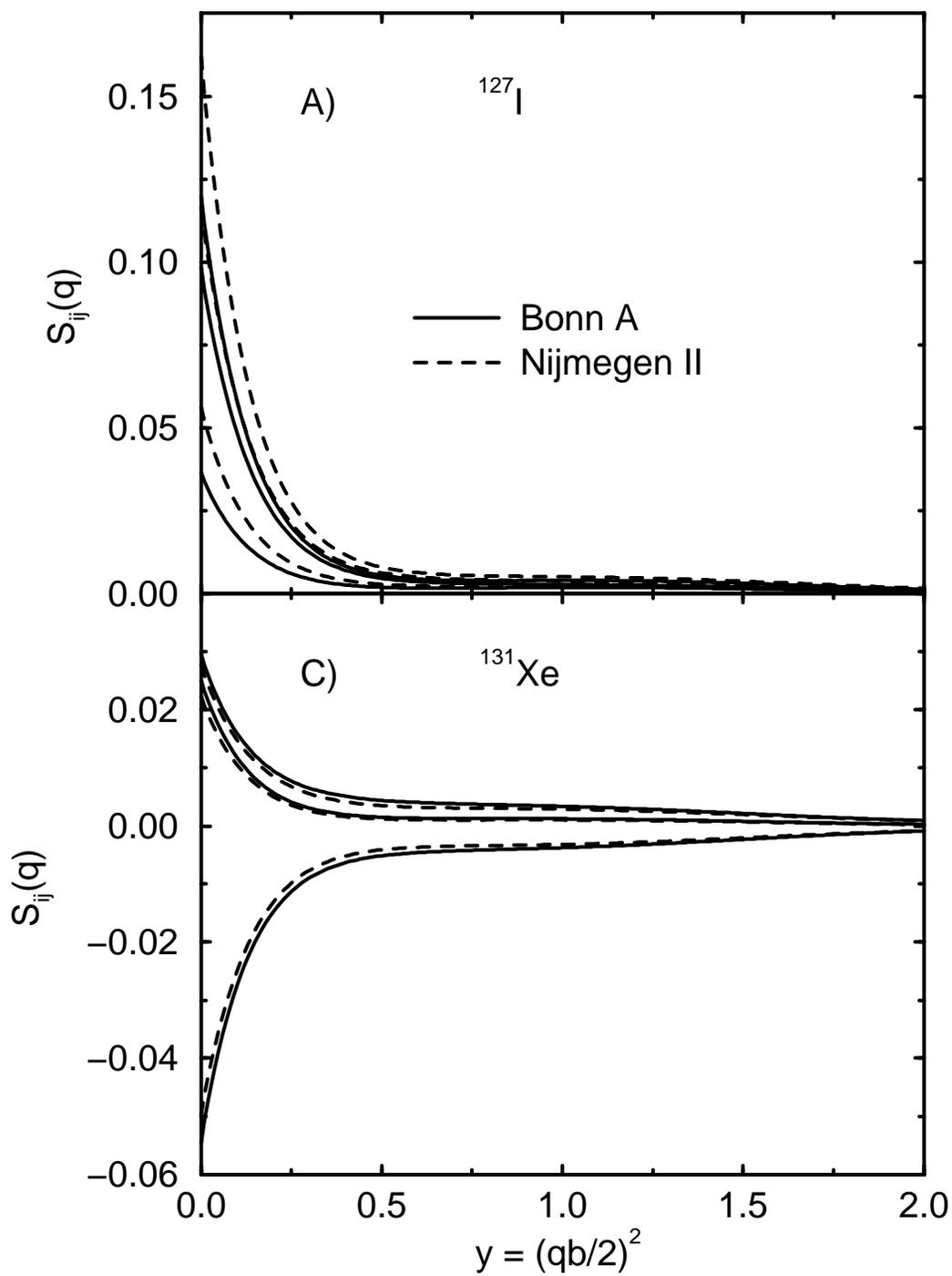
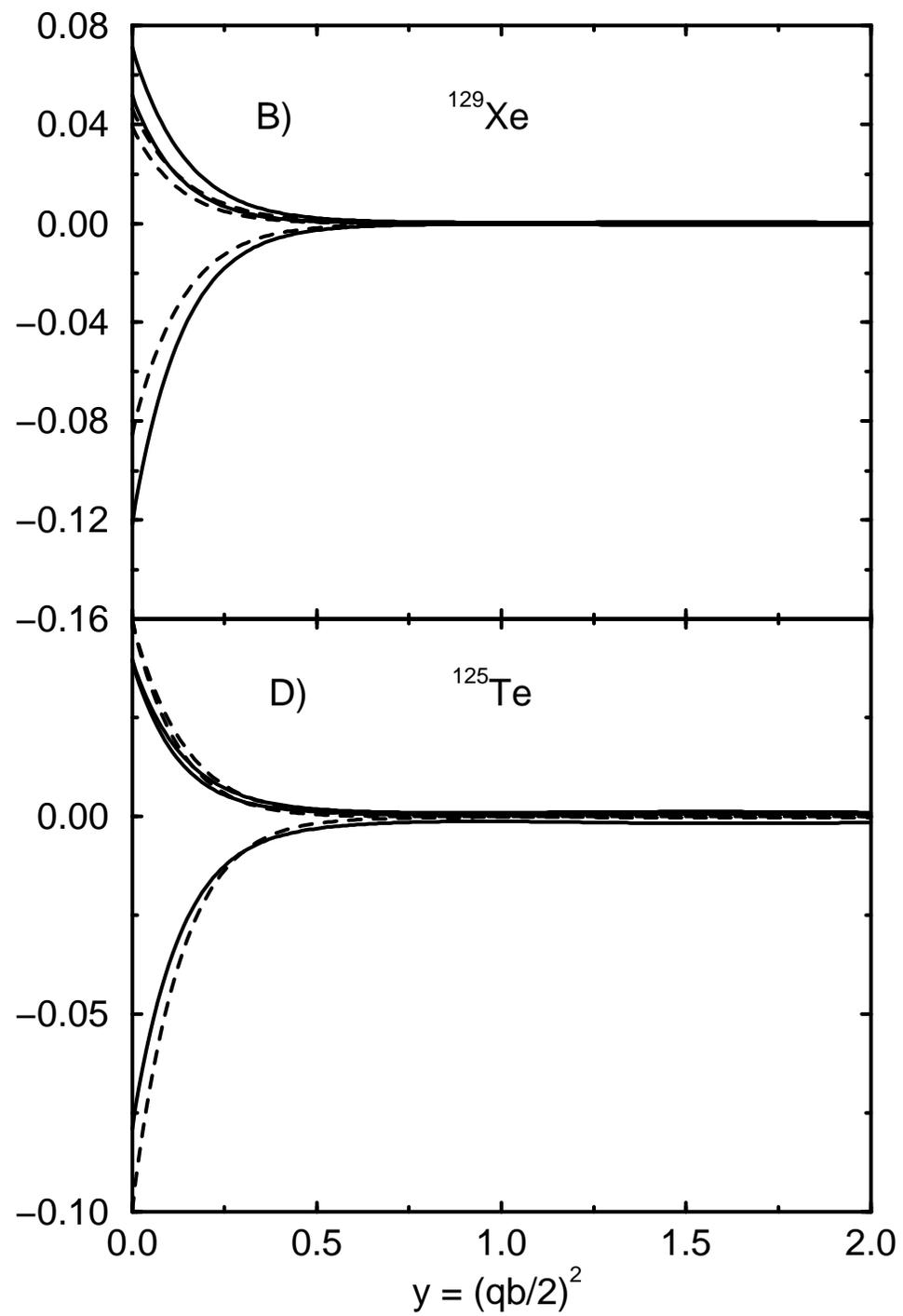
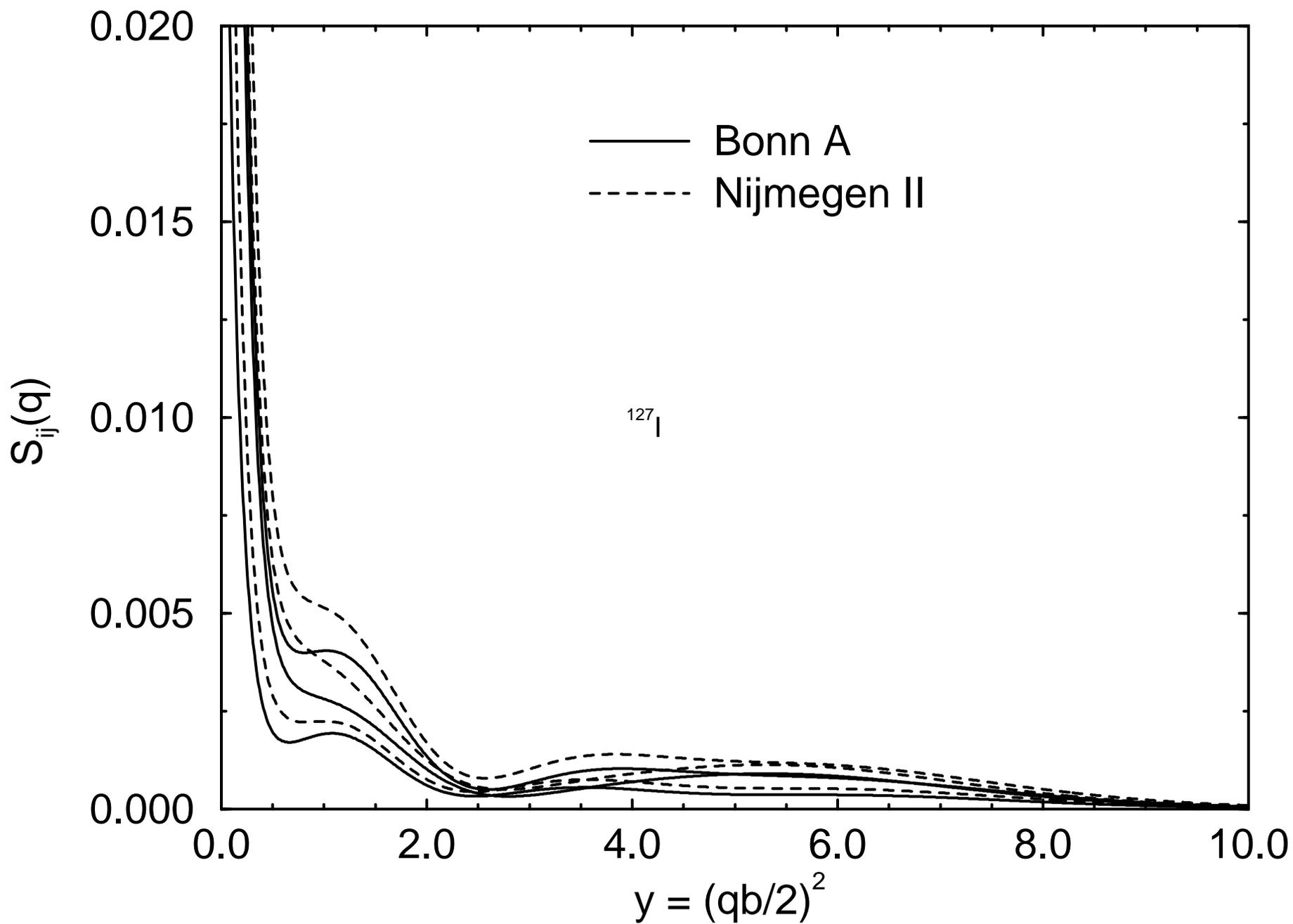
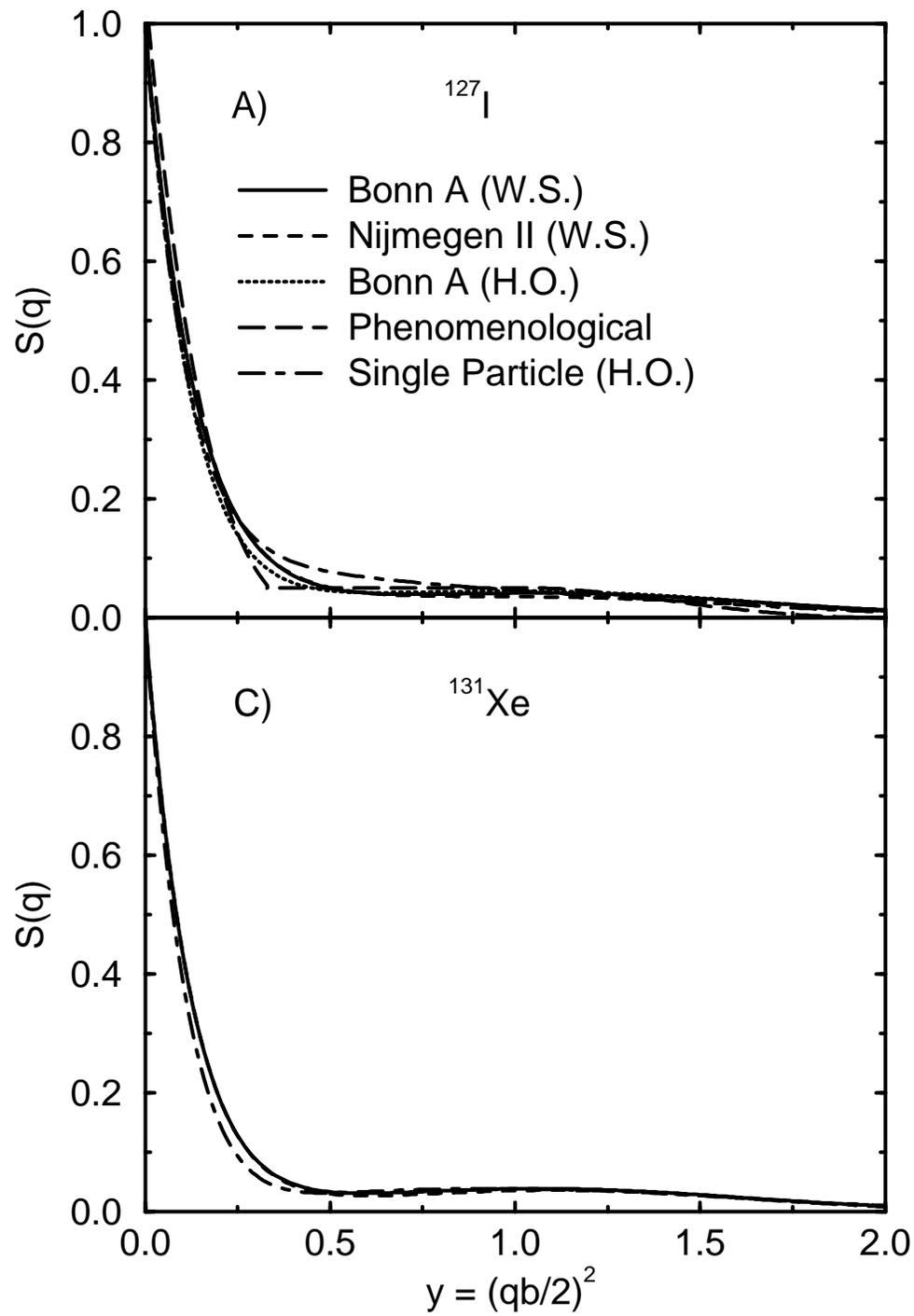
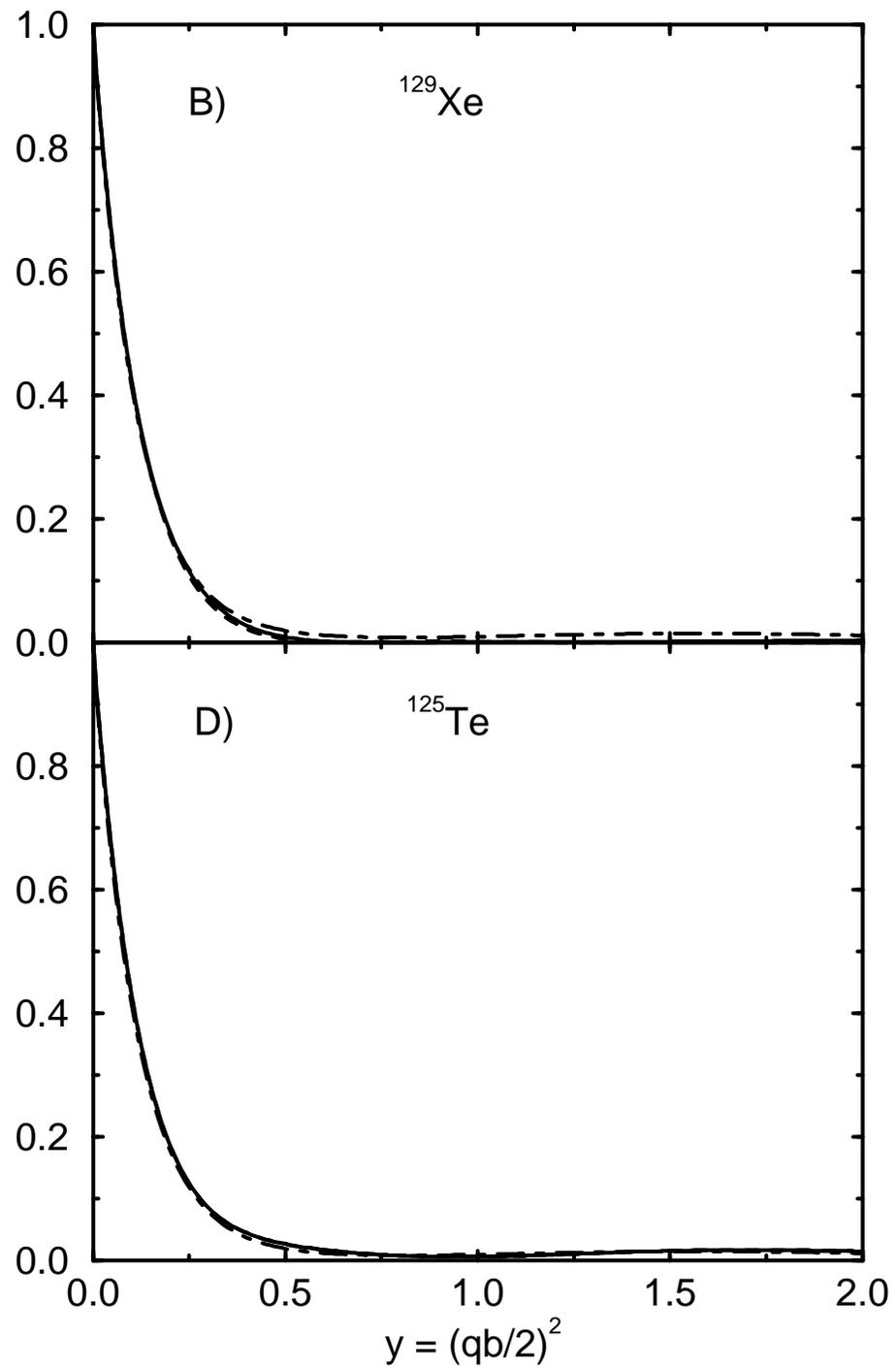
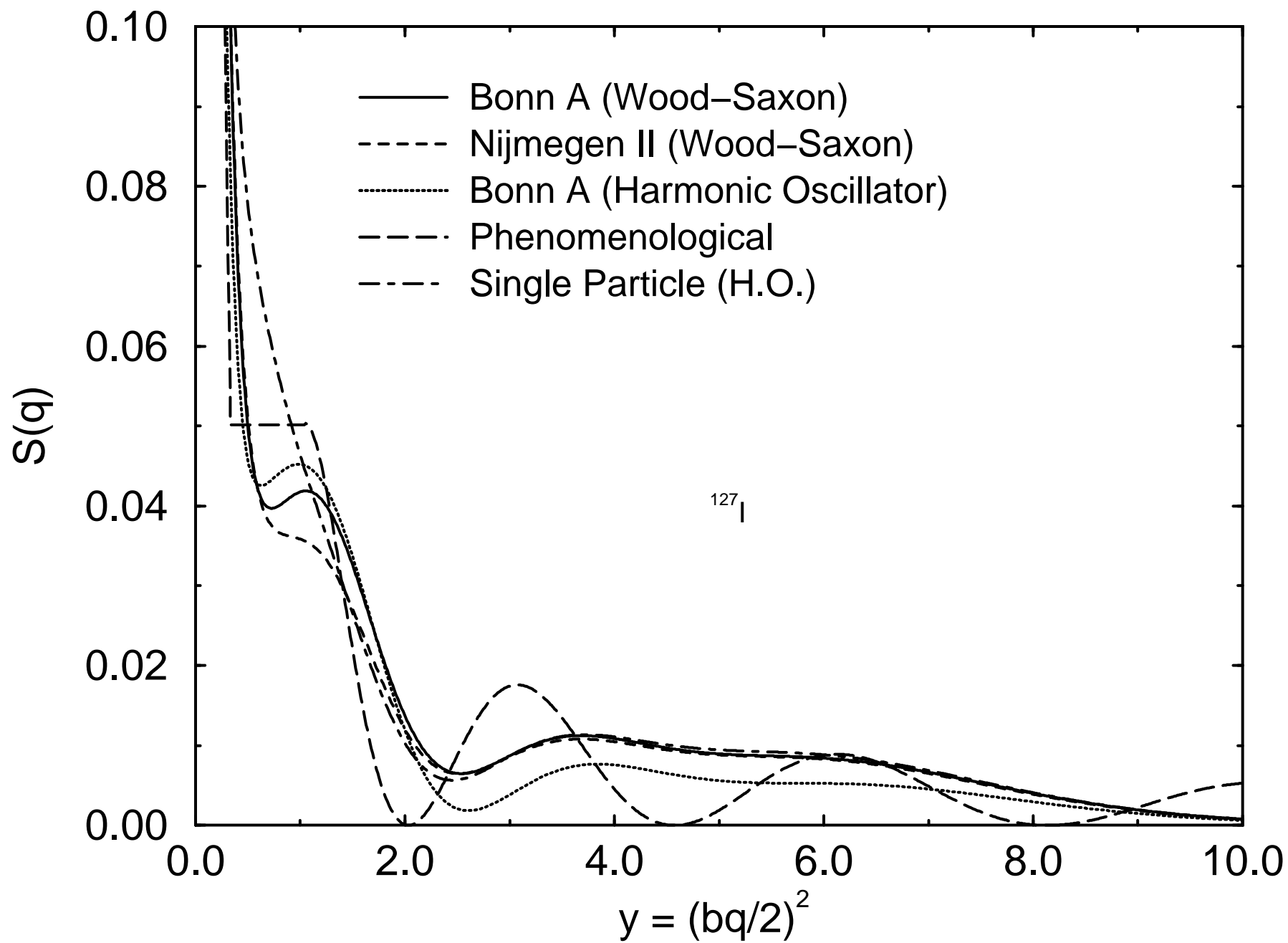
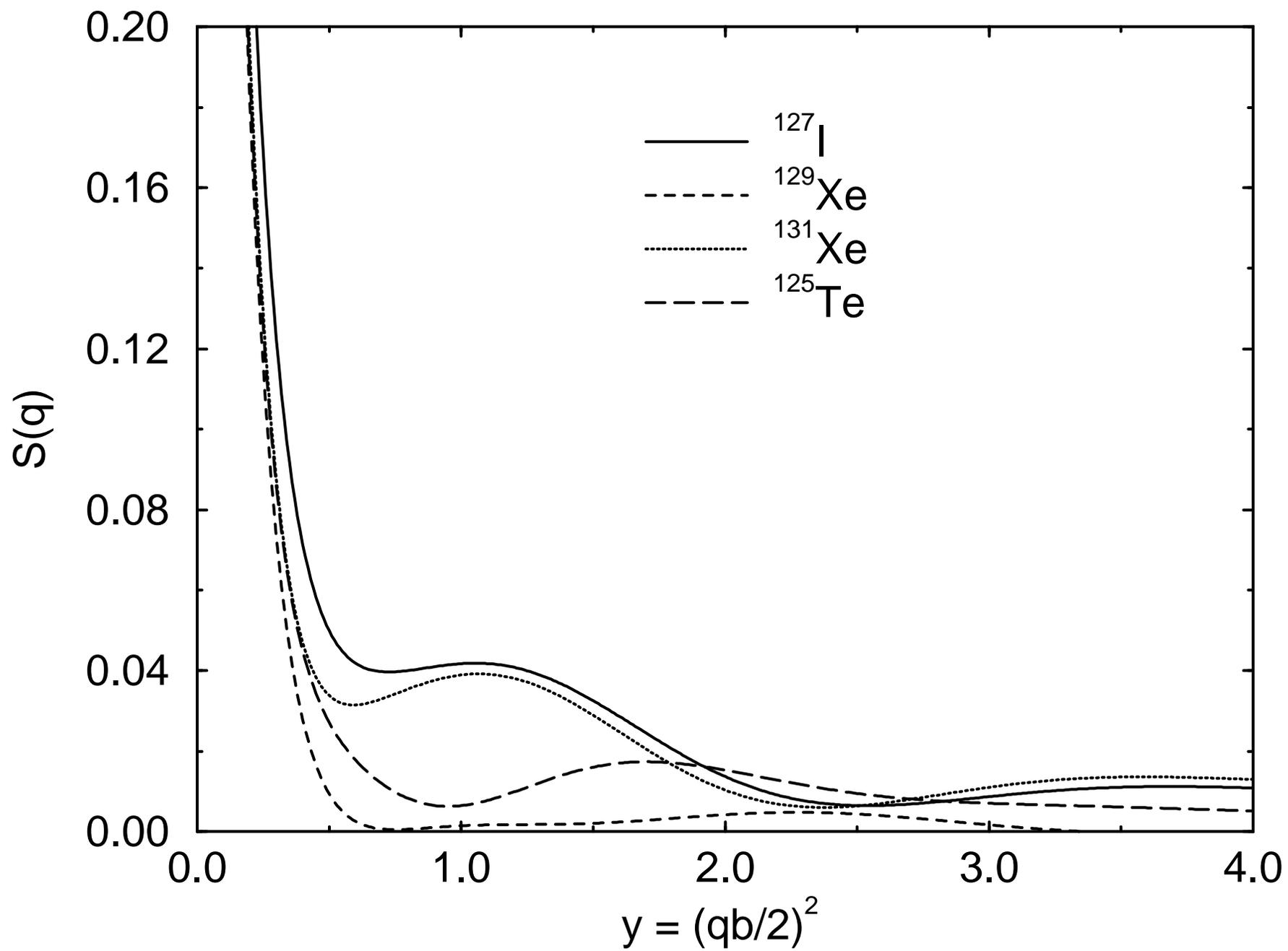
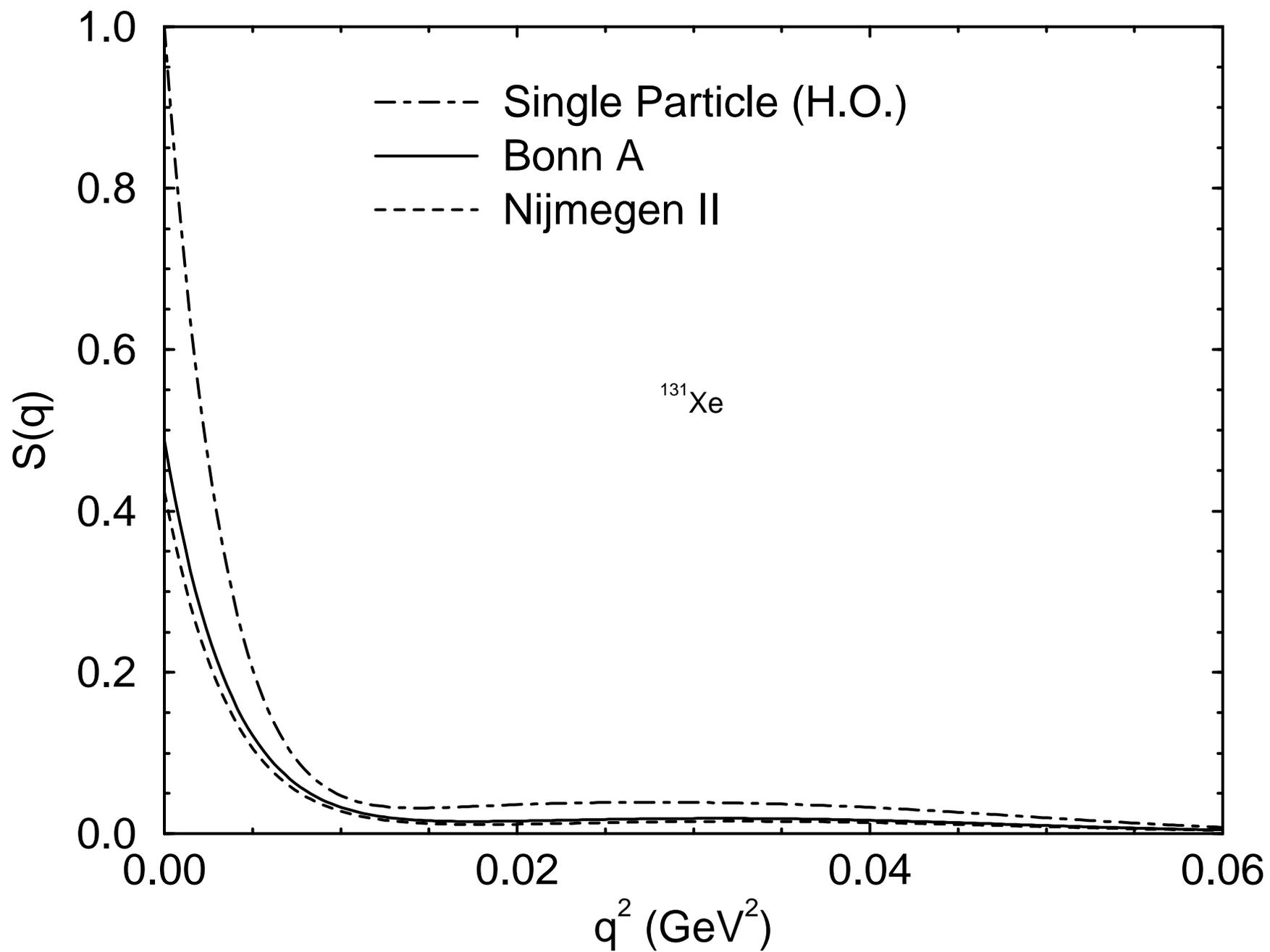
\end{document}